\documentclass{icrc}

\usepackage{times}
\usepackage{graphicx} % when using Latex and dvips
\begin{document}

\title{Neutrino Emission from HBLs and LBLs}
\author[1]{A. M\"ucke\footnote{Now at: 
Institut f\"ur Theoretische Physik, Lehrstuhl
IV: Weltraum- \& Astrophysik, Ruhr-Universit\"at Bochum, D-44780 Bochum,
Germany}}
\affil[1]{D\'epartement de Physique,
Universit\'e de Montr\'eal, Montr\'eal, Qu\'ebec, H3C 3J7, Canada}
 
\author[2]{R.J. Protheroe}
 
\affil[2]{Department of Physics and
Mathematical Physics, The University of Adelaide, Adelaide, SA 5005,
Australia}
 
\correspondence{muecke@astro.umontreal.ca (Email)}

\firstpage{1}
\pubyear{2001}

% \titleheight{11cm} % uncomment and adjust in case your title block
                     % does not fit into the default and minimum 7.5 cm

\maketitle

\begin{abstract}
The Synchrotron Proton Blazar model is a promising model to
explain high energy emission from $\gamma$-ray loud BL Lac
objects like Mkn~421. In contrast to leptonic models, the
hadronic explanation of $\gamma$-ray emission predicts ultrahigh
energy neutrinos.

The predicted neutrino spectra from a typical High-energy cutoff BL Lac
Object (HBL) and a Low-energy cutoff BL Lac Object (LBL) are
presented. We find that cooling due to muon synchrotron radiation
causes a cutoff of the neutrino spectrum at $\sim 10^{18}$ eV,
with the exception of $\nu_\mu$ from kaon decay which may extend
to higher energies if meson production takes place in the
secondary resonance region of the cross section.

The impact of the neutrino output from both source populations to
the diffuse neutrino background is discussed.
\end{abstract}

\section{Introduction}

Most models of gamma ray emission from blazar jets consider
Inverse Compton scattering of energetic electrons off
low energy photons as the main gamma ray production mechanism. In
the Synchrotron-Proton Blazar (SPB) model, recently proposed by,
e.g., \citet{MP01} accelerated protons interact with the
synchrotron radiation field produced by the co-accelerated
electrons via meson photoproduction and Bethe-Heitler pair
production, and more importantly, with the strong ambient
magnetic field, emitting synchrotron radiation (pions and muons
also emit synchrotron radiation).  Since this model neglects
external photon field components, it is most applicable to BL~Lac
objects which have weak accretion disks. \citet{MP01} have shown
that this model can reproduce the commonly observed double-humped
blazar spectral energy distribution (SED).

Gamma ray loud BL~Lac objects are commonly classified as HBLs or
LBLs on the basis of their ratio of radio to X-ray flux i.e. HBLs
have a broad-band spectral index $\alpha_{\rm{RX}} \leq 0.75$ and
LBLs have $\alpha_{\rm{RX}} > 0.75$ \citep{PG95}.  Consequently,
for LBLs, the synchrotron peak is generally observed at
optical/IR-frequencies while the X-ray band covers the local
minimum of $\nu L_{\nu}$ between the two spectral humps. LBLs are
thought to represent an intermediate object class between
quasars, which have generally a much higher bolometric luminosity
than BL~Lac objects, such that $\nu L_{\nu}$ peaks in the IR and
at MeV-GeV-energies, and HBLs, which are the least luminous
blazars, and have $\nu L_{\nu}$ spectral peaks at soft to
medium-energy X-rays and GeV-TeV gamma rays.

In this paper we present for the first time the predicted
neutrino emission from a typical HBL and LBL in
the SPB model, assuming that Mkn~421 and PKS~0716+714 are typical
for each source population, respectively.

\section{The model}

We inject an spectrum of $E_p^{'-2}$ protons into the jet (primed
quantities are in the jet frame), and they are assumed to remain
quasi-isotropic in the jet frame due to pitch-angle scattering.
The co-accelerated relativistic electrons radiate synchrotron
photons which serve as the target radiation field for
proton-photon interactions, and for the subsequent
pair-synchrotron cascade which develops as a result of
photon-photon pair production in the highly magnetized
environment. In the SPB model, the target photon density
$u'_{\rm{phot}}$ is much smaller than the magnetic field energy
density, and thus Inverse Compton losses can be neglected. 
The pair-synchrotron
cascade redistributes the photon
 power to lower energies where the photons
escape from the
 emission region, or ``blob'', of size $R'$ which moves
relativistically in a direction closely aligned with our
line-of-sight.

The proton energy loss processes considered in the model are
photomeson production, Bethe Heitler pair production, proton
synchrotron radiation and adiabatic losses due to jet expansion.
Synchrotron radiation from $\mu^\pm$ and $\pi^\pm$ (from photomeson
production) prior to their decay becomes important in highly
magnetized environments with $B'>$ several tens of Gauss
\citep{RM98}, and is taken into account in our calculations.
The acceleration rate for any acceleration mechanism is $dE'/dt'
= \eta(E') e c^2 B'$ where $\eta(E') \le 1$ describes the
efficiency, and $B'$ is the magnetic field in the blob.  The
maximum proton energy is limited by the balance between energy
gain and loss rates.

Direct proton and muon synchrotron radiation seems to be mainly
responsible for the high energy hump whereas the low energy hump
is dominanted by synchrotron radiation from the directly
accelerated $e^-$, with a contribution of synchrotron radiation
from secondary electrons produced by the proton- and
$\mu^\pm$-synchrotron initiated cascade.  The contribution from
Bethe-Heitler pair production turned out to be negligible.  For
our calculations we use a Monte-Carlo method and utilize the
recently developed SOPHIA code for the photohadronic event
generation \citep{SOPHIA}.

\section{HBLs and LBLs in the SPB Model: the
case Mkn~421 and PKS~0716+714}

Mkn~421 is a well-known TeV-blazar, and is classified as an HBL.
We have used the average SED published by \citet{G98} to 
derive its model parameters in the framework of our model.

The {\it{observed}} low energy hump, identified as synchrotron
radiation, serves as the target photon field in our model for
pion photoproduction and cascading. We have parametrized the
observations by a broken power law with photon spectral index
--1.5 below the lab frame break energy $\epsilon_b \approx 100$
eV, and index --2.25 up to the cutoff energy of
$\epsilon_c\approx 40$ keV.  

Our model represents the data well \citep{MP02} for Doppler
factor $D\approx 10$, $\epsilon_b' \approx 10$ eV, and
$\epsilon_c' \approx 4$ keV, u'$_{\rm{phot}} \approx 10^{9}$ eV/cm$^3$,
an emitting volume of $\approx 10^9$AU$^3$, $B'\approx 50$G,
ambient proton energy density $u'_P \approx 10^2$ erg cm$^{-3}$
and $\eta\approx 0.8$ .  Fig.~1 shows the relevant time scales.
The clear dominance of the proton synchrotron radiation at high
energies is apparent, while pion production is rather of minor
importance. The proton spectrum is cut off at $\gamma_p'\approx
10^{10.5}$ due to proton synchrotron radiation.  Significant
$\pi^\pm$ synchrotron losses do not occur below the proton
cutoff, while $\mu^\pm$ synchrotron radiation cannot be
neglected.

\begin{figure}[t] 
\vspace*{2.0mm} % just in case for shifting the figure slightly down 
\includegraphics[width=8.3cm]{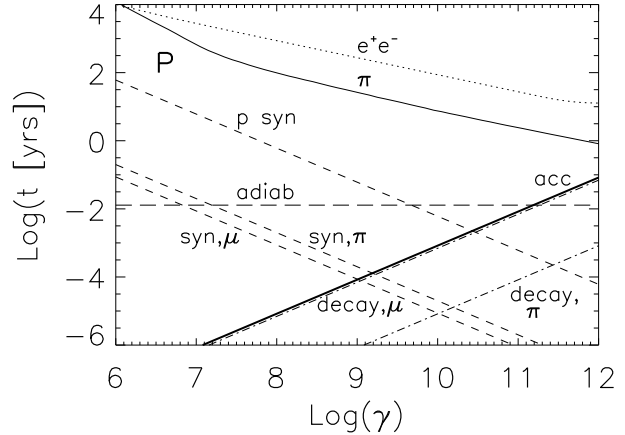}  
\caption{Mean (jet frame) energy loss and acceleration time
scales for the HBL Mkn~421 as modeled in \citet{MP02}:
$\pi$-photoproduction ($\pi$), Bethe-Heitler pair production
($e^+e^-$) and proton synchrotron radiation (syn); $\pi^\pm$- and
$\mu^\pm$ for synchrotron radiation (syn $\pi$, syn $\mu$) are
also shown and compared with their mean decay time scales (decay
$\pi$, decay $\mu$). The acceleration time scale (acc) is
indicated as a thick straight line.  Model parameters are:
$B'=50$ G, $D = 10$, $R'\approx 10^{16}$ cm, $u'_{\rm{phot}} =
2\times 10^{9}$ eV cm$^{-3}$, $u'_p \approx 100$ erg cm$^{-3}$,
$\gamma'_{p\rm{,max}} = 3 \times 10^{10}$, $\eta = 0.8$.  }
\end{figure}

The LBL PKS~0716+714 has been chosen for its well-defined
low-energy synchrotron component \citep{G98}. Again, the same
power-law spectral indices has been used as before to represent
this component, a lab-frame break energy $\epsilon_b \approx
0.7$ eV and cutoff energy $\epsilon_c \approx 3.5$ keV.  Our
model parameters used are: $u'_{\rm{phot}} \approx 10^{11}$
eV/cm$^3$, $D\approx 7$, $R' \approx 10^{17}$cm, $B'\approx 30$G,
$u'_p \approx 40$ erg cm$^{-3}$ and $\eta\approx 10^{-2}$.

In this environment, losses due to photo-meson production limit
the injection proton spectrum to about
$\gamma_{\rm{p,max}}\approx 10^9$ (see Fig.~2). With proton
synchrotron radiation being rather unimportant here, photo-pion
production dominates the electromagnetic energy output, and is
responsible for a stronger neutrino flux than in the HBL case
(see Sect.~4).  Due to the low cutoff of the proton spectrum,
pion synchrotron radiation can again be neglected.

\begin{figure}[t] 
\vspace*{2.0mm} % just in case for shifting the figure slightly down 
\includegraphics[width=8.3cm]{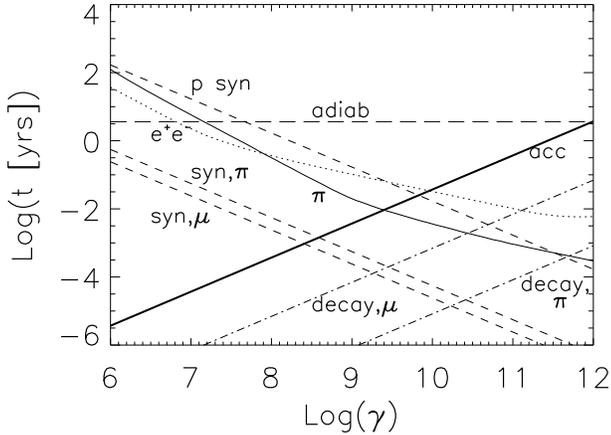}
\caption{ Mean (jet frame) energy loss and acceleration time
scales for the LBL PKS~0716+714  as modeled in
\citet{MP02}. Annotations are the same as in Fig.~1.  Model
parameters are: $B'=30$ G, $D\approx 7$, $R'\approx 10^{17}$ cm,
$u'_{\rm{phot}} = 4\times 10^{11}$ eV/cm$^3$, $u'_p = 40$
erg/cm$^{-3}$, $\gamma'_{\rm{p,max}} = 3 \times 10^{9}$, $\eta \approx
10^{-2}$.}
\end{figure}

Our modeling seems to favour the hypothesis that the low
luminosities observed from HBLs might be due to an
{\it{intrinsically}} low jet-frame target photon density while
the more luminous LBLs may possess rather {\it{intrinsically}}
high jet-frame target photon fields.  The consequences of this
hypothesis for the Synchrotron-Proton Blazar model will be
presented elsewhere \citep{MP02}.

\section{Neutrino Spectra}

Hadronic blazar models produce neutrino emission mainly through
the production and decay of charged mesons,
e.g. $\pi^\pm\to\mu^\pm +\nu_{\mu}/\bar{\nu_{\mu}}$ followed by
$\mu^\pm\to e^\pm + \nu_e/\bar{\nu_e} +
\nu_{\mu}/\bar{\nu}_{\mu}$.  The neutrinos escape without further
interaction.

Fig.~3 shows the predicted neutrino output from Mkn~421, which is
considered as a typical HBL, and the typical LBL PKS~0716+714. We
give the predicted neutrino emission from the objects themselves,
and do not consider here any additional contribution from
escaping cosmic rays interacting while propagating through the
cosmic microwave background radiation.

The clear dominance of neutrino emission from LBLs in comparison
to HBLs is obvious. The reason for this is the higher meson
production rate in the LBL source population (see also Fig.~1 and
2) due to their higher target photon fields in comparison to the
HBL population.
 
The proton injection spectrum is modified by photohadronic
losses.  For Mkn~421 the photopion production rate approximately
follows a broken power law, $t_\pi^{-1} \propto \gamma_p^{1.25}$ for
proton energies below $\sim 10^{7}$~GeV, and $\propto
\gamma_p^{0.5}$ above $E_p' \sim 10^{7}$~GeV (see Fig.~1) due to
the break in the target photon spectrum. This leads to a break in
the neutrino spectrum at $\sim 10^{7}$~GeV (observer frame), from
power spectral index
$\alpha_\nu \approx 1.25$ to $\alpha_\nu \approx 0.5$ 
($(E^2 dN/dE) \propto E^{\alpha_\nu}$).  The
cutoff at $\sim 10^9$ GeV in the observer's frame is caused by
$\mu^\pm$-synchrotron losses, whereas $\pi^\pm$-synchrotron
emission turns out to be unimportant.  Another important source
of high energy neutrinos is the production and decay of charged
kaons if the proton-photon interaction takes place predominantly
in the secondary resonance region of the cross section
\citep{SOPHIA}. This might be the case for HBLs because their
target photon field can extend up to X-ray energies.  The
positively charged kaons decay in $\sim 64\%$ of all cases into
muons and direct high energy muon-neutrinos. In contrast to the
neutrinos originating from $\pi^\pm$ and $\mu^\pm$-decay, these
muon-neutrinos will not have suffered energy losses through
$\pi^\pm$- and $\mu^\pm$-synchrotron radiation, and therefore
appear as an excess in comparison to the remaining neutrino
flavors at the high energy end ($E_\nu \approx
10^{9}$--$10^{10}$~GeV) of the emerging neutrino spectrum, and in
addition cause the total neutrino spectrum to extend to $\sim
10^{10}$~GeV.

In contrast, the neutrino emission from PKS~0716+714 is cutoff at
$\sim 10^9$ GeV (observer frame) for all neutrino flavors (see
Fig.~3) due to a roughly one order of magnitude lower proton
cutoff. Also $\mu^\pm$ synchrotron losses may play a role
here. The neutrino spectrum follows a power law with index
$\alpha_\nu \approx 1.25$ below the cutoff, and is caused by
photohadronic interactions with the target photon field above
$\epsilon_b'$. Because of the $\pi$-production threshold and the
relatively low proton cutoff in LBLs, meson production in the
photon field below $\epsilon_b'$ cannot occur, leading to a
simple power law for the neutrino spectrum.

The photon-hadron interactions for both, LBLs and HBLs, take
place predominantly in the resonance region. Here, $\pi^-$, and
thus $\bar{\nu_e}$ production is suppressed (see Fig.~3).
 
Previous hadronic jet models expected equal photon and neutrino
energy fluxes (e.g. \citet{Ma93}).  Our model predicts for HBLs a
peak neutrino energy flux approximately two to three orders of
magnitude lower than high energy gamma rays ($q_{\gamma\nu}
\approx 10^{2\ldots 3}$), while for LBLs the neutrino and gamma
ray output is approximately comparable ($q_{\gamma\nu} \approx
1$).

 \begin{figure*}[t] 
 \begin{center}
 \includegraphics*[width=9.5cm]{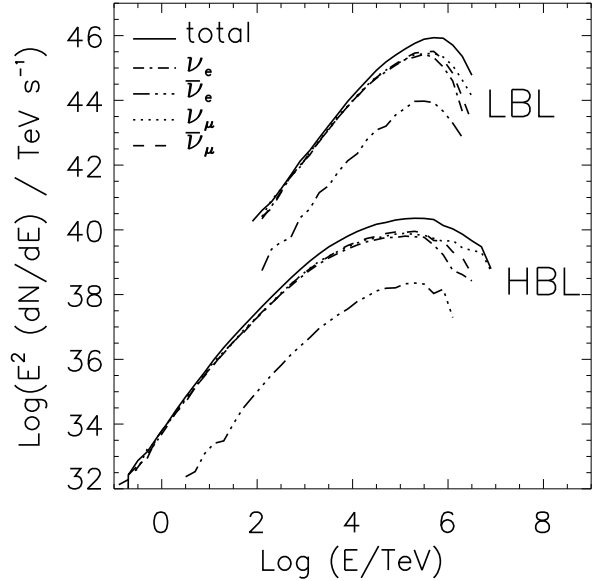}
 \begin{minipage}{9.5cm}
 \caption{Predicted neutrino spectra for the HBL Mkn~421 (lower
 curves) and the LBL PKS~0716+714 (upper curves) as modeled in \citet{MP02}.} 
 \end{minipage}
 \end{center}
 \end{figure*}

\section{Diffuse neutrino background}

From inspecting Fig.~3 it is immediately clear that LBLs would
dominate the BL Lac contribution to the diffuse neutrino
background unless HBLs turn out to be significantly more numerous
than LBLs.

A commonly used method to estimate the diffuse neutrino
background due to unresolved blazars is to determine the ratio
gamma ray to neutrino emission, $q_{\gamma\nu}$, in the objects
in question, and then normalize to the observed extragalactic
gamma ray background (e.g. \citet{Ma95}).  The different photon
to neutrino ratio we found in HBLs and LBLs leads to various
implications for the diffuse neutrino emission.

The extragalactic diffuse gamma ray background (EGRB) as detected
by EGRET is $(1.45 \pm 0.05) \times 10^{-5}$ photons cm$^{-2}$
s$^{-1}$ sr$^{-1}$ above 100 MeV \citep{kumar98}. Thus, the
extragalactic neutrino background (ENB) due to unresolved EGRET
blazars can be estimated by: $(E^2 dN/dE)_{\rm{ENB}} \approx 7.4
(1+f) q_s q_{\gamma\nu}^{-1} \times 10^{-7}$ GeV cm$^{-2}$
s$^{-1}$ sr$^{-1}$ where $f<1$ is the fraction of photon power
lost in the cosmic background matter and radiation field, and
$q_s\approx 0.25$ is the presumed blazar contribution to the
EGRET EGRB \citep{CM98}. 
Propagation of gamma rays at $\leq 100$ GeV is 
relatively unattenuated, and is therefore comparable to neutrino 
propagation, so we set $f=0$.
Since LBLs are EGRET-sources, and with
$q_{\gamma\nu} \approx 1$, a diffuse neutrino flux of
$(E^2 dN/dE)_{\rm{ENB}} \approx 2\times 10^{-7}$ GeV
cm$^{-2}$ s$^{-1}$ sr$^{-1}$ follows {\it{if LBLs dominate the
blazar contribution to the EGRET diffuse gamma ray background}}.
This is slightly above the neutrino upper bound derived using
cosmic ray constraints by \citet{WB99} and \citet{MPR01} for
optically thin sources, but well below the upper bound for
optically thick sources. Thus, we conclude that either LBLs do
not dominate the blazar contribution to the EGRET EGRB, or LBLs
are rather thick sources (like quasars).

Our estimate of the ENB due to HBLs follows a similar
procedure. Since HBLs emit, however, most of their photon power
at TeV-energies, we extrapolate the $E^{-2.1}$ EGRB spectrum as
observed by EGRET up to 25 TeV, the highest energy observed from
any HBL, to estimate the unattenuated EGRB flux in the $\geq 100$
GeV energy range. The resulting ENB from HBLs is
then
$(E^2 dN/dE)_{\rm{ENB}} \approx 3 q_s \times
(10^{-9}\ldots 10^{-10})$ GeV cm$^{-2}$ s$^{-1}$ sr$^{-1}$, where
$q_s$ is here the HBL contribution to our estimated diffuse
TeV-background. Even for $q_s=1$ the predicted ENB intensity
lies several orders of magnitude below the neutrino upper bounds
(Waxman \& Bahcall 1999; Mannheim et al.\ 2001, MPR), and is also
lower than the estimate from MPR of the BL Lac contribution to
the ENB using the observed XBL luminosity function.  Note,
however, that MPR assumed pion photoproduction was the main
proton loss process, whereas in the SPB model for HBLs
proton synchrotron losses dominate at the expense of pion (and neutrino)
production.  The same is true for \citet{ICRC99}, where one to
two order of magnitude higher contribution from TeV-blazars to
the diffuse neutrino background was estimated in comparison to
this work.

\section{Summary}

Neutrino production spectra of BL Lac objects are calculated in
the framework of the Synchrotron-Proton Blazar model \citep{MP01}
using the SOPHIA event generator for photomeson production
\citep{SOPHIA}.  The $\nu$ spectra presented here for Mkn~421
and PKS~0716+714, typical of HBLs and LBLs, respectively, are
modeled using a parameter set which reproduces well the average
SED of those blazars. We find higher pion production rates for
LBLs in comparison to the HBLs, and this leads a higher neutrino
output for a typical LBL by $\sim 10^6$ compared to HBLs. Muon
synchrotron losses are responsible for a spectral cutoff in the
neutrino SED at $\sim 10^{18}$ eV, with a possible extension of
$\nu_\mu$s up to $\sim 10^{19}$ eV, due to kaon decay.

The total power deposited in gamma rays and neutrinos is
approximately equal for LBLs, while a two to three order of
magnitude smaller neutrino peak flux is predicted in HBLs in
comparison to their peak photon power.  This leads to a
significantly higher LBL contribution to the extragalactic
diffuse neutrino background than from HBLs, when the method of
normalizing to the observed EGRB is used for estimating the
diffuse neutrino flux.

\begin{acknowledgements}
AM acknowledges a postdoctoral bursary from the Qu\'ebec
Government.  The research of RJP is funded by a grant from the
Australian research Council.
\end{acknowledgements}

\end{document}